\documentclass{aa}  

\usepackage{graphicx}
\usepackage{txfonts}
\begin{document}

   \title{A search for Galactic transients disguised as gamma--ray bursts}

%   \subtitle{}

   \author{A.~Clerici\thanks{aurora.clerici@student.unife.it}
          \inst{1}
          \and
          C. Guidorzi\inst{1}%\thanks{guidorzi@fe.infn.it}%\fnmsep\thanks{}
          \and
          V. La~Parola\inst{2}
          }

   \institute{Department of Physics and Earth Sciences, University of Ferrara, via Saragat 1,
                 I-44122, Ferrara, Italy\\
              \and
              INAF -- Istituto di Astrofisica Spaziale e Fisica Cosmica di Palermo, via U. La~Malfa 153,
                 I-90146, Palermo, Italy\\
%              \email{adriano.baldeschi@student.unife.it}
%         \and
%             University of Alexandria, Department of Geography, ...\\
%             \email{c.ptolemy@hipparch.uheaven.space}
%             \thanks{The university of heaven temporarily does not
%                     accept e-mails}
             }

%   \date{Received September 15, 1996; accepted March 16, 1997}

% \abstract{}{}{}{}{} 
% 5 {} token are mandatory: context, aims, methods, results, conclusions
 
   \abstract
       {A significant fraction of cosmological gamma--ray bursts (GRBs) are characterised by a fast rise
         and exponential decay (FRED) temporal structure. This is not a distinctive feature of this class, since
         it is observed in many Galactic transients and is likely descriptive of a sudden release of energy followed
         by a diffusion process. Possible evidence has recently been reported by \citet{Tello12} for a Galactic
         contamination in the sample of FRED GRBs discovered with {\em Swift}.}
       {We searched for possible Galactic intruders disguised as FRED GRBs in the {\em Swift} catalogue up to
         September 2014.}
       {We selected 181 FRED GRBs (2/3 with unknown redshift) and considered different subsamples.
         We tested the degree of isotropy through the dipole and the quadrupole moment distributions, both with
         reference to the Galaxy and in a coordinate-system---independent way, as well as with the
         two--point angular autocovariance function.
         In addition, we searched for possible indicators of a Galactic origin among the spectral and temporal
         properties of individual GRBs.}
       {We found marginal ($\sim$$3\sigma$) evidence for an excess of FREDs with unknown redshift towards the
         Galactic plane compared with what is expected for an isotropic distribution corrected for the non-uniform
         sky exposure. However, when we account for the observational bias against optical follow-up
           observations of low-Galactic latitude GRBs, the evidence for anisotropy decreases to $\sim$$2\sigma$.
           In addition, we found no statistical evidence for different spectral or temporal properties from
           the bulk of cosmological GRBs.}
         { We found marginal evidence for the presence of a disguised Galactic population among {\em Swift} GRBs
             with unknown redshift. The estimated fraction is $f=(19\pm11)$\%, with an upper limit of 34\%
             (90\% confidence).}

   \maketitle
%
%________________________________________________________________

%%%%%%%%%%%%%%%%%%%%%%%%%%%%%%%%%%%%%%%%%%%%%%
\section{Introduction}
\label{sec:intro}
%%%%%%%%%%%%%%%%%%%%%%%%%%%%%%%%%%%%%%%%%%%%%%
Gamma--ray bursts (GRBs) are known to be cosmological transient sources that signal the
very final stage of some kind of massive stars for the long duration ones, and likely
the act of merging for a compact binary system for short duration ones
(see \citealt{MeszarosGehrels12} for a review). Owing to their extreme luminosity
they are routinely detected from cosmological distances up to redshifts $z\sim9.4$
\citep{Cucchiara11b}. The consequent highly isotropic sky distribution
was interpreted as evidence for a cosmological origin prior to the first redshift
measurements \citep{HartmannEpstein89,HartmannBlumenthal89,Briggs96}.
Once the cosmological nature of long-duration GRBs was finally established, a number
of studies reported evidence for anisotropy in the distribution of short-duration ones
(\citealt{Balazs98,Magliocchetti03,Vavrek08}; but see also \citealt{Bernui08}), or
evidence for a correlation between short GRBs and galaxies in the local Universe
\citep{Tanvir05}.

Recently, Tello~et~al. \nocite{Tello12} (2012; hereafter T12) exploited the real-time
arcmin-sized localisation capabilities of the {\em Swift} Burst Alert Telescope
(BAT; \citealt{Barthelmy05}) to test the isotropy of a special class of long GRBs,
the so-called fast-rise exponential-decay (FRED) ones. Not only is their time profile
quite common in GRBs \citep{Norris96}, but it is also descriptive of various high-energy
outbursts from Galactic sources, such as X--ray binaries \citep{Remillard06rev},
X--ray bursters \citep{Lewin93rev}, magnetars \citep{Mereghetti08rev},
as well as unidentified Galactic transients \citep{Kasliwal08,CastroTirado08,Stefanescu08}.
Such universality is observed across different wavelengths all the way to
the radio bands, where analogous flares are observed over timescales
from seconds or less, up to years, due to a broad range of astrophysical sources \citep{Pietka15}.
The ubiquity of FRED-like transient profiles is explained by its being descriptive
of a sudden energy release, followed by some cooling and diffusion process.
T12 found some evidence ($<3 \sigma$ confidence) for a deviation from isotropy of
the sky distribution of {\em Swift} FREDs with unknown $z$. They attribute it
to the presence of Galactic unidentified sources which would make up to $\sim27$\%
of the observed unknown-$z$ FRED population.

Motivated by their results, we carried out a similar investigation on an updated
sample of {\em Swift} FRED catalogue, adopting independent selection criteria.
We analysed the degree of isotropy both with reference to the Galactic plane and
in a reference-system--independent way. In addition, to further characterise the
nature of the Galactic candidates, we examined on a statistical basis the spectral
and temporal properties of the $\gamma$--ray prompt and X--ray afterglow emission
as observed with {\em Swift} BAT and X--ray Telescope (XRT; \citealt{Burrows05})
and compared with a sample of cosmological GRBs.
The paper is organised as follows: the sample selection and data analysis are
described in Sections~\ref{sec:data_sel} and \ref{sec:data}, respectively.
Section~\ref{sec:res} presents the results, which are discussed in Section~\ref{sec:disc}.

%%%%%%%%%%%%%%%%%%%%%%%%%%%%%%%%%%%%%%%%%%%%%%%%%%%%%%%%%%%
\section{Sample selection}
\label{sec:data_sel}
\begin{figure*}
%\begin{center}
\sidecaption
\includegraphics[width=12.5cm]{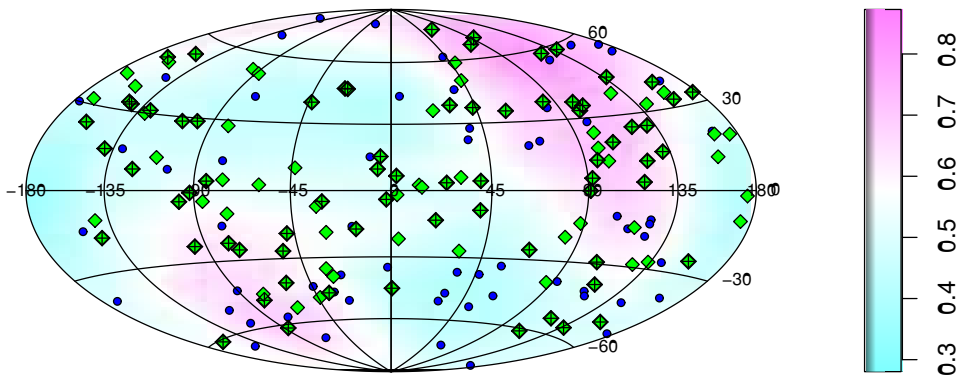}
\caption{Exposure map of BAT in Galactic coordinates from January 2005 to September 2014. Green diamonds
(blue circles) are all FREDs with unknown (known) $z$. They jointly constitute S1.
All green diamonds are sample S2. Crossed green diamonds are those unknown-$z$ FREDs with a unique
detected peak (sample S3). The three sets include 181, 119, and 71 GRBs, respectively.}
\label{fig:RITD}
%\end{center}
\end{figure*}

%%%%%%%%%%%%%%%%%%%%%%%%%%%%%%%%%%%%%%%%%%%%%%%%%%%%%%%%%%%
Starting from a sample of 904~GRBs detected by {\em Swift}-BAT in the time interval January 2005
to September 2014, we selected those which had explicitly been tagged by the BAT team as FREDs
in the GCN circulars. We ended up with 181 FREDs out of 904~GRBs, whereas T12 found 111 FREDs out of 596 GRBs.
We further characterised the number of peaks of each GRB by means of the
MEPSA code \citep{Guidorzi15a}, capable of identifying relatively faint peaks at very different timescales.
We set a threshold of $4.5$ on the minimum signal--to--noise as yielded by MEPSA, which ensures a negligible
false positive rate for our light curve sample. 
We then considered three different samples: i) the full one; ii) the subsample of GRBs with unknown
redshift $z$; iii) the subsample of GRBs with unknown $z$ and with a single peak detected with MEPSA.
Hereafter these samples are referred to as S1, S2, and S3, including 181, 119, and 71 GRBs, respectively.
Their sky distributions in Galactic coordinates are shown in Fig.~\ref{fig:RITD}.
To assess the role of observational biases, for comparison we also consider two additional samples detected
in the same time interval: (i) all 904 BAT-discovered GRBs, and (ii) the complementary sample of 506
non--FREDs GRBs with unknown $z$. Hereafter, we refer to them as STot and SnFnZ, respectively.
Table~\ref{tab:sizes} summarises the samples' sizes and the corresponding values of T12.
\begin{table}  
  \begin{center}         
    \caption{GRB samples' sizes.}
    \label{tab:sizes} 
    \begin{tabular}{llrr}
       \hline 
      \noalign{\smallskip} 
      Sample  & Description & Size &  Size\\
              &             & (this work) & (T12)\\
      \noalign{\smallskip}
      \hline
       \noalign{\smallskip} 
      S1    &  all FREDs                 & 181  &  111\\
      S2    &  all FREDs, no $z$         & 119  &   77\\
      S3    &  all FREDs, no $z$, 1 peak &  71  &   49\\
      STot  &  all GRBs                  & 904  &   -\\
      SnFnZ &  all non-FREDs, no $z$     & 506  &   -\\
       \noalign{\smallskip}
      \hline
    \end{tabular}
  \end{center}
%  \tablefoot{
%    \tablefoottext{a}{Estimated through MC simulations (method 1).}
%  } % end tablefoot
\end{table} 
%

%%%%%%%%%%%%%%%%%%%%%%%%%%%%%%%%%%%%%%%%%%%%%%%%%%%
\section{Data analysis}
\label{sec:data}
%%%%%%%%%%%%%%%%%%%%%%%%%%%%%%%%%%%%%%%%%%%%%%%%%%%
For each real sample we calculated the three Galactic dipole moments ($\sin{b}$, $\cos{b}\sin{l}$, $\cos{b}\cos{l}$),
and the five quadrupole moments ($\sin^2{b}-1/3$, $\cos^2{b}\sin{2l}$, $\cos^2{b}\cos{2l}$, $\sin{2b}\sin{l}$,
$\sin{2b}\cos{l}$) with an emphasis on the ones related to the Galactic plane, i.e. $\sin{b}$ and
$(\sin^2{b}-1/3)$. We also calculated the somewhat redundant equatorial dipole $\sin{\delta}$ and
quadrupole $(\sin^2{\delta}-1/3)$ moments to better evaluate possible anisotropies connected with
observational biases.
In addition, we calculated the two--point angular correlation function \citep{HartmannBlumenthal89}.
We also carried out a reference-system--independent search for anisotropies possibly unrelated with our Galaxy
by evaluating the Rayleigh--Watson ${\mathcal W}$ and the Bingham ${\mathcal B}$ statistics \citep{Briggs93}.
Everything was then compared with the expectations for isotropy following two independent approaches:
through Monte Carlo simulations and sky pixelisation.

The exposure map of {\em Swift}-BAT (Fig.~\ref{fig:RITD}) was obtained from the limiting flux map for the
considered time interval, processing the BAT data with the Bat\_Imager software \citep{Segreto10}, and
following the procedure described in \citet{Cusumano10}.

%-----------------------------------------------
\subsection{Monte Carlo simulations (method 1)}
\label{sec:simul}
%-----------------------------------------------
For each sample we generated $10^3$ synthetic samples, each having as many positions as the corresponding
real one. Each fake position was generated according to the following procedure: i) a random position
is drawn from an isotropic distribution; ii) the position is accepted/rejected depending on the outcome of a
binomial variate, whose acceptance probability was set to the fractional exposure of that position.
The procedure ends as soon as the desired number of positions is achieved.

%-----------------------------------------------
\subsection{Sky pixelisation (method 2)}
\label{sec:pix}
%-----------------------------------------------
The entire sky was split into a homogeneous grid of 5292 pixels following the technique described
by \citet{Tegmark96}, with angular resolution of $3\fdg2$. Each pixel was then assigned the corresponding
fractional exposure value. The choice of the angular resolution was driven by the need to ensure
an adequate coverage of the BAT exposure map.
The expected moments were calculated over the entire sky grid by weighting the contribution of
each sky pixel by the associated fractional exposure.
The uncertainties affecting the observed moments were calculated from the expected variances
in the Gaussian limit ensured by the central limit theorem due to $N\gg1$ events:
$\sigma^2(\langle \sin{b}\rangle) = 1/(3N)$, $\sigma^2(\langle \sin^2{b}\rangle) = 4/(45N)$
for the $(\sin^2{b}-1/3)$ term, $\sigma^2 = 4/(15N)$ for the other
quadrupole terms \citep{Briggs93}.

\setcounter{table}{1}  %%%%%%%%%%%%%%%%%%%%%%%%%%%%%%%%%%%%%%%%%%%% TABLE 2
 \begin{table*}  
 \tabcolsep 4pt         
 \begin{center}         
 \caption{Average dipolar and quadrupolar moments for the different samples.
   The real value moments are reported and associated to the values obtained from the sky pixelisation method,
   along with their discrepancies in $\sigma$ units. The last two columns report the MC $1\sigma$ and $2\sigma$
   (Gaussian--equivalent) confidence intervals.}
 \label{tab:moments}       
 \begin{tabular}{lcrrccc}
 \hline 
 \hline 
\noalign{\smallskip} 
  Sample  & Moment &  Observed & Expected\tablefootmark{(a)}  & Discrepancy & \multicolumn{2}{c}{Expected Interval (MC)\tablefootmark{(b)}} \\
          &        &           & (sky pix.; $\pm1\sigma$) & ($\sigma$)  & ($1\sigma$)& ($2\sigma$)\\
 \noalign{\smallskip}
 \hline
 \noalign{\smallskip} 
S1 & $\langle \cos{b}\cos{l}\rangle$  & $-0.010$ & $-0.010\pm0.043$ & $0.0$ & $-0.050, +0.031$ &  $-0.091, +0.072$\\
S1 & $\langle \sin{b}\rangle$         &  $0.028$ &  $0.022\pm0.043$ & $+0.1$ & $-0.018, +0.064$ &  $-0.061, +0.105$\\
S1 & $\langle \sin{\delta}\rangle$    & $ 0.065$ & $0.035\pm0.043$ & $+0.7$ & $-0.012, +0.074$ &  $-0.052, +0.120$\\
S1 & $\langle \sin^2{b} -1/3\rangle$  & $-0.009$ &  $0.014\pm0.022$ & $-1.0$ & $-0.009, +0.035$ &  $-0.030, +0.057$\\
S1 & $\langle\sin^2{\delta}-1/3\rangle$&$ 0.074$ & $ 0.039\pm0.022$ & $+1.5$ & $+0.016, +0.062$ &  $-0.005, +0.085$\\
  \hline
S2 & $\langle \cos{b}\cos{l}\rangle$  & $-0.005$ & $-0.010\pm0.053$ & $+0.1$ & $-0.063, +0.036$ &  $-0.108, +0.081$\\
S2 & $\langle \sin{b}\rangle$         &  $0.097$ &  $0.022\pm0.053$ & $+1.4$ & $-0.030, +0.075$ &  $-0.089, +0.125$\\
S2 & $\langle \sin{\delta}\rangle$    & $ 0.070$ & $0.035\pm0.053$ & $+0.7$ & $-0.018, +0.086$ &  $-0.072, +0.141$\\
S2 & $\langle \sin^2{b} -1/3\rangle$  & $-0.063$ &  $0.014\pm0.027$ & $-2.8$ & $-0.012, +0.043$ &  $-0.036, +0.072$\\
S2 & $\langle\sin^2{\delta}-1/3\rangle$&$ 0.103$ & $ 0.039\pm0.027$ & $+2.3$ & $+0.011, +0.068$ &  $-0.016, +0.098$\\
  \hline
S2\tablefootmark{(c)} & $\langle \cos{b}\cos{l}\rangle$  & $-0.005$ & $0.017\pm0.053$ & $-0.4$ & $-0.036, +0.066$ &  $-0.080, +0.123$\\
S2\tablefootmark{(c)} & $\langle \sin{b}\rangle$         &  $0.097$ &  $-0.008\pm0.053$ & $+2.0$ & $-0.057, +0.043$ &  $-0.108, +0.096$\\
S2\tablefootmark{(c)} & $\langle \sin{\delta}\rangle$    & $ 0.070$ & $-0.022\pm0.053$ & $+1.7$ & $-0.079, +0.034$ &  $-0.130, +0.085$\\
S2\tablefootmark{(c)} & $\langle \sin^2{b} -1/3\rangle$  & $-0.063$ &  $-0.013\pm0.027$ & $-1.8$ & $-0.042, +0.015$ &  $-0.069, +0.045$\\
S2\tablefootmark{(c)} & $\langle\sin^2{\delta}-1/3\rangle$&$ 0.103$ & $ 0.056\pm0.027$ & $+1.7$ & $+0.031, +0.086$ &  $-0.001, +0.114$\\
  \hline
S3 & $\langle \cos{b}\cos{l}\rangle$  & $-0.005$ & $-0.010\pm0.068$ & $+0.1$ & $-0.080, +0.059$ &  $-0.148, +0.117$\\
S3 & $\langle \sin{b}\rangle$         &  $0.111$ &  $0.022\pm0.068$ & $+1.3$ & $-0.046, +0.088$ &  $-0.109, +0.165$\\
S3 & $\langle \sin{\delta}\rangle$    & $ 0.097$ & $0.035\pm0.068$ & $+0.9$ & $-0.038, +0.106$ &  $-0.109, +0.171$\\
S3 & $\langle \sin^2{b} -1/3\rangle$  & $-0.032$ &  $0.014\pm0.035$ & $-1.3$ & $-0.025, +0.048$ &  $-0.059, +0.083$\\
S3 & $\langle\sin^2{\delta}-1/3\rangle$&$ 0.102$ & $ 0.039\pm0.035$ & $+1.8$ & $+0.004, +0.078$ &  $-0.030, +0.121$\\
  \hline
SnFnZ & $\langle \cos{b}\cos{l}\rangle$  & $ 0.036$ & $-0.010\pm0.026$ & $+1.8$ & $-$ &  $-$\\
SnFnZ & $\langle \sin{b}\rangle$         &  $0.011$ &  $0.022\pm0.026$ & $ 0.0$ & $-$ &  $-$\\
SnFnZ & $\langle \sin{\delta}\rangle$    & $ 0.040$ & $0.035\pm0.026$ & $+2.0$ & $-$ &  $-$\\
SnFnZ & $\langle \sin^2{b} -1/3\rangle$  & $-0.014$ &  $0.014\pm0.013$ & $-2.1$ & $-$ &  $-$\\
SnFnZ & $\langle\sin^2{\delta}-1/3\rangle$&$ 0.046$ & $ 0.039\pm0.013$ & $+0.5$ & $-$ &  $-$\\
  \hline
STot & $\langle \cos{b}\cos{l}\rangle$  & $ 0.014$ & $-0.010\pm0.019$ & $+1.2$ & $-$ &  $-$\\
STot & $\langle \sin{b}\rangle$         &  $0.011$ &  $0.022\pm0.019$ & $-0.6$ & $-$ &  $-$\\
STot & $\langle \sin{\delta}\rangle$    & $ 0.040$ & $0.035\pm0.019$ & $+0.2$ & $-$ &  $-$\\
STot & $\langle \sin^2{b} -1/3\rangle$  & $ 0.017$ &  $0.014\pm0.010$ & $+0.4$ & $-$ &  $-$\\
STot & $\langle\sin^2{\delta}-1/3\rangle$&$ 0.037$ & $ 0.039\pm0.010$ & $-0.2$ & $-$ &  $-$\\
  \noalign{\smallskip}
  \hline
  \end{tabular}
  \end{center}
\tablefoot{
\tablefoottext{a}{Estimated through the sky pixelisation (method 2) in the Gaussian limit ensured by the central limit theorem.}
\tablefoottext{b}{Estimated through MC simulations (method 1).}
\tablefoottext{c}{S2 is compared with expectations that modelled observational biases connected with the redshift measurement (Section~\ref{sec:bias}).}
 } % end tablefoot
  \end{table*}

%%%%%%%%%%%%%%%%%%%%%%%%%%%%%%%%%%%%%%%%%%%%%%%%%%%
\section{Results}
\label{sec:res}
%%%%%%%%%%%%%%%%%%%%%%%%%%%%%%%%%%%%%%%%%%%%%%%%%%%
%-----------------------------------------------
\subsection{Dipole and quadrupole moments}
\label{sec:dqmom}
%-----------------------------------------------
%
\begin{figure*}
\begin{center}
\includegraphics[width=18cm]{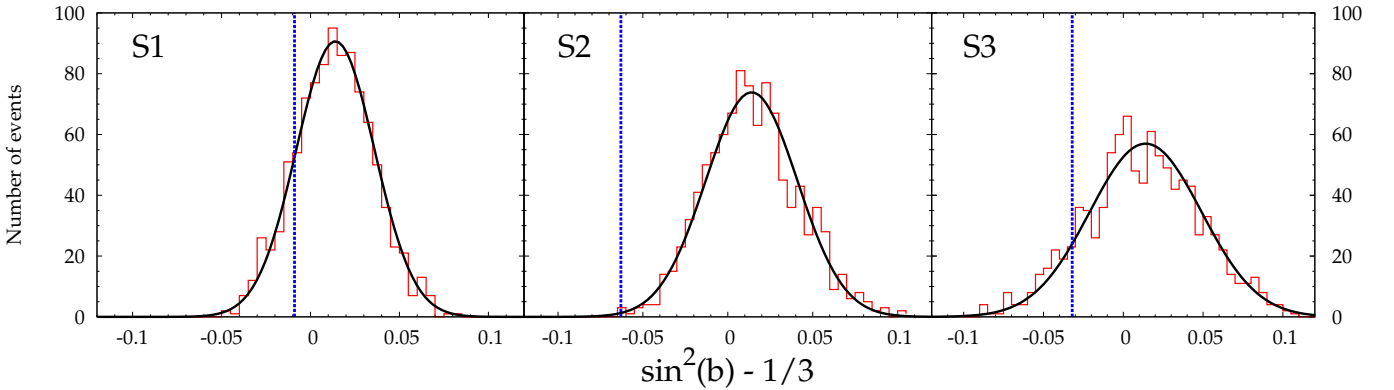}
\caption{Galactic quadrupole moment distributions obtained from MC simulations (method 1; histograms)
and from the sky pixelisation in the Gaussian limit (method 2; solid curves)
compared with the values observed in the real samples (vertical lines) for S1, S2, and S3 (left,
mid, and right panels, respectively).}
\label{fig:moments}
\end{center}
\end{figure*} 
Table~\ref{tab:moments} reports the observed moments along with the expected values and intervals.
While none of the dipole moments deviate by $>$$2 \sigma$ from isotropy, the Galactic
quadrupole moment of S2 is marginally ($-2.8 \sigma$) lower than the isotropic--expected one.
Analogous result is obtained for SnFnZ with a less significant deviation though ($-2.1 \sigma$).
This suggests that the unknown-$z$ FREDs tend to cluster at low Galactic latitudes, or equivalently,
that they are rarely found at high latitudes.
Among the other Galactic quadrupole moments, only $\sin{2b}\,\cos{l}$ exhibits two $>2$$\sigma$ deviations
from isotropy, $-2.5 \sigma$ ($-2.4 \sigma$) for S1 (S2). 
Figure~\ref{fig:moments} shows the Galactic quadrupole moment $(\sin^2{b}-1/3)$ for each of the three
samples obtained with both methods assuming isotropy, corrected for the non-uniform exposure map,
to be compared with the value observed in the corresponding real sample.

%-----------------------------------------------
\subsection{${\mathcal W}$ and ${\mathcal B}$ statistics}
\label{sec:WBstat}
%-----------------------------------------------
Following \citet{Briggs93}, for each position $(l_i,b_i)$ we considered the associated unit
vector $\vec{r}_i=(x_i,y_i,z_i)$ and calculated the Rayleigh-Watson statistic defined as
\begin{equation}
{\mathcal W} = \frac{3}{N}\ \Big|\sum_{i=1}^N \vec{r}_i\Big|^2\;.
\label{eq:W}
\end{equation}
This statistic characterises the dipole moment and in the absence of any preferred direction
is asymptotically distributed as $\chi^2_3$ for large $N$. The Bingham statistic ${\mathcal B}$ is defined as
\begin{equation}
{\mathcal B} = \frac{15\,N}{2}\ \ \sum_{k=1}^3 \Big(\lambda_k - \frac{1}{3}\Big)^2\;,
\label{eq:B}
\end{equation}
where $\lambda_k$ are the eigenvalues of the so-called ``orientation'' matrix, that is equivalent
to the quadrupole one and is defined as
\begin{equation}
M_N = \frac{1}{N} \sum_{i=1}^N \ \left(\begin{array}{ccc}
x_i x_i & x_i y_i & x_i z_i\\
y_i x_i & y_i y_i & y_i z_i\\
z_i x_i & z_i y_i & z_i z_i \end{array} \right)\;.
\label{eq:MN}
\end{equation}
${\mathcal B}$ measures the deviation of the eigenvalues from the value $1/3$ expected for
isotropy and is asymptotically distributed as $\chi^2_5$. 
Because of the non-uniform sky exposure we applied just method~1 to derive the expected intervals
for both statistics.
Table~\ref{tab:WBstat} reports the observed values and the corresponding intervals for isotropy.
Like for the Galactic moments we did not find evidence for any preferred direction with a non-zero
dipole moment, all within $1 \sigma$ confidence. As for the quadrupole term, although all of them
lie within $2 \sigma$, S2 shows the largest deviation. In particular, it mildly suggests a clustering
of unknown-$z$ FREDs along the Galactic direction $l=130^\circ$, $b=+13^\circ$ ($\alpha=39^\circ$,
$\delta=+75^\circ$, J2000) and its antipodal one, i.e. close to celestial poles.
From Table~\ref{tab:moments} the equatorial quadrupole moment $(\sin^2{\delta}-1/3)$ exceeds by $2.3 \sigma$ the
expected isotropic value only for S2, thus lending support to the enhanced presence of unknown-$z$
FREDs towards the celestial poles.
\setcounter{table}{2}  %%%%%%%%%%%%%%%%%%%%%%%%%%%%%%%%%%%%%%%%%%%% TABLE 3
 \begin{table}  
 \tabcolsep 4pt         
 \begin{center}         
 \caption{Observed values of Rayleigh--Watson (${\mathcal W}$) and Bingham (${\mathcal B}$) statistics
and confidence intervals expected for an isotropic distribution corrected for the non--uniform exposure.}
 \label{tab:WBstat}       
 \begin{tabular}{lccrrcrr}
 \hline 
 \hline 
\noalign{\smallskip} 
  Sample  & Statistics &  Observed & \multicolumn{5}{c}{Expected Interval (MC)\tablefootmark{a}} \\
          &        &               & \multicolumn{2}{c}{($1\sigma$)} &  & \multicolumn{2}{c}{($2\sigma$)}\\
%          &        &               & \multicolumn{2}{r}         & \multicolumn{2}{r}\\
 \noalign{\smallskip}
 \hline
 \noalign{\smallskip} 
S1 & ${\mathcal W}$ & $\ 2.6$ & $0.9$ & $6.0$ & & $0.2$ & $10.5$\\
S2 & ${\mathcal W}$ & $\ 3.7$ & $0.9$ & $5.6$ & & $0.2$ & $10.2$\\
S3 & ${\mathcal W}$ & $\ 3.3$ & $0.9$ & $5.5$ &  &$0.2$ & $10.1$\\
  \hline
S1 & ${\mathcal B}$ & $14.2$ & $4.9$ & $16.0$ & & $1.7$ & $24.8$\\
S2 & ${\mathcal B}$ & $19.4$ & $3.7$ & $13.8$ & & $1.6$ & $21.5$\\
S3 & ${\mathcal B}$ & $\ 9.8$ & $2.9$ & $11.1$ &  &$1.1$ & $18.6$\\
  \hline
  \noalign{\smallskip}
  \hline
  \end{tabular}
  \end{center}
\tablefoot{
\tablefoottext{a}{Estimated through MC simulations (method 1).}
 } % end tablefoot
  \end{table} 

The unknown-$z$ non-FREDs GRBs (SnFnZ) show a similar behaviour: their Galactic quadrupole moment
deviates by $-2.1 \sigma$ from isotropy. In contrast, in spite of the best statistical sensitivity
the whole BAT--detected sample (STot) shows no evidence at all for non-zero quadrupole moments.
%%%%%%%% The following sentence is no more true, after finding a mistake in expected sin(delta). %%%%%%%%
%whereas a $+2.67 \sigma$ significant dipole moment is found towards the celestial North Pole.

%-----------------------------------------------
\subsection{Two-point angular autocovariance}
\label{sec:2pAC}
\begin{figure}
\includegraphics[width=8.5cm]{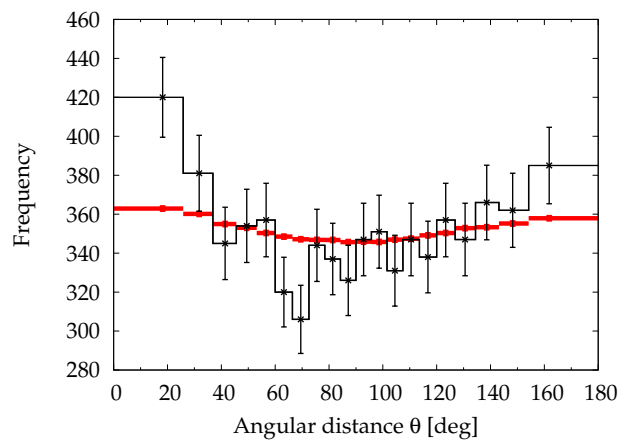}
\caption{Histogram of the angular autocovariance function distribution observed for S2 compared
with what is expected taking into account the non-uniform sky exposure (thick red line). The bin width
is uniform in $\cos{\theta}$.}
\label{fig:PADs2}
\end{figure}

%-----------------------------------------------
Finally, we calculated the distribution of the two-point angular autocovariance function by
taking all the pairs within a given sample. In particular, we considered $\cos{\theta}$ rather
than $\theta$ (angular distance between a pair), given that for isotropy and uniform exposure
the expected distribution is uniform in $\cos{\theta}$. Likewise, the expected distribution for
isotropy was derived only through method 1 to account for the non-uniform exposure map.
The most remarkable deviations are observed for S2, as shown in Fig.~\ref{fig:PADs2}.

The thick line shows the isotropic-expected value as a function of the angular distance. The displayed
uncertainties on the observed numbers are their square root values, under the assumption of Poisson
statistics. Strictly speaking, this is not correct: the ensemble of $\theta$ values are not
fully independent of each other, since the entire set of possible pairs is not.
However, given the large number of pairs the degree of correlation is relatively small, so as a
matter of fact Poissonian uncertainties are realistic.

Unknown-$z$ FREDs seem to cluster on angular scales $<30^\circ$ (between $2 \sigma$ and $3 \sigma$
significance) at the expense of those in the range $60^\circ$--$90^\circ$. A $\chi^2$ test yields
a p--value of 15\% ($\chi^2/{\rm dof}=25.4/19$). However, this must be taken as a loose indication,
for the lack of statistical independence of the different bins.

%

%-----------------------------------------------
\subsection{Observational biases connected with redshift measurement}
\label{sec:bias}
\begin{figure}
\includegraphics[width=8.5cm]{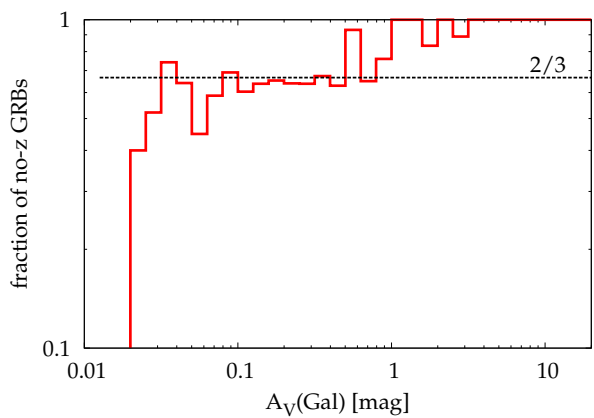}
\caption{Fraction of GRBs with unknown redshift as a function of Galactic dust extinction.}
\label{fig:AVdist}
\end{figure}

%-----------------------------------------------
In the light of the previous results, we tried to account for the observational biases
involved in measuring $z$ that come into play when one considers the
samples of unknown-$z$ GRBs with some evidence for anisotropy, S2 and SnFnZ.
To this aim, using method~1 we generated another set of 1000~synthetic samples with
the same size as S2, which took into account two main sources of bias:
\begin{itemize}
\item Galactic dust extinction disfavours optical observations of low--Galactic-latitude
GRBs;
\item ground-based follow-up observations are less probable around the celestial poles
\citep{Fynbo09}.
\end{itemize}
To model the first bias, we considered the Galactic extinction as measured by \citet{Schlafly11}
in terms of $A_V$.\footnote{Available at \url{http://irsa.ipac.caltech.edu/applications/DUST/}}
We then obtained the $A_V({\rm Gal})$ distribution for the positions of all the GRBs
belonging to STot and split it into two groups, with and without measured $z$ respectively.
For each $A_V({\rm Gal})$ bin we calculated the fraction of GRBs with unknown $z$ over the total.
The result is shown in Fig.~\ref{fig:AVdist}.
As long as it is $A_V({\rm Gal})<1$~mag the fraction is about $2/3$, i.e. the same value that is
obtained over the entire {\em Swift} catalogue. However, for $A_V({\rm Gal})>1$~mag almost
all of the GRBs have no redshift. 
In generating the synthetic samples analogous to S2, each random position was accepted depending
on the outcome of a binomial variate: the acceptance probability was set to 1 if $A_V({\rm Gal})>1$~mag,
to $2/3$ otherwise.

To model the effect of the second bias, we preliminarily compiled a list of the ground facilities
that mostly contributed to measuring GRB redshifts. For each declination, we considered the visibility
from each location by calculating the fraction of time that a generic source at the given declination
spends at $>+30^\circ$ above the local horizon. Such threshold value matches average observational
constraints. We then averaged it over the full set of locations, weighting by the multiplicity of
the number of telescopes at a given location (e.g., in La Palma there are several telescopes which
provided GRB redshift measurements). We have therefore come up with a probability of having $z$ measured
as a function of declination. This bias was incorporated in the MC simulations as a further criterion
for a random position to be accepted, depending on the outcome of a binomial variate: the acceptance
probability was set to the complement to one of the probability of having $z$ measured for that declination.

We also applied method~2 by weighting each sky pixel by the product of the sky exposure and the
two above probabilities. We then calculated the mean value for each moment and the corresponding
uncertainties as in Section~\ref{sec:pix} under the same assumptions.
The results are reported in Table~\ref{tab:moments}.
Both methods gave fully consistent results as was the case for the other samples.

It is noteworthy that the deviation from isotropy of the Galactic quadrupole moment $(\sin^2{b}-1/3)$
tapers off from $-2.8 \sigma$ to $-1.8 \sigma$ when observational biases are accounted for.
Likewise, the equatorial quadrupole moment $(\sin^2{\delta}-1/3)$ decreases from $+2.3 \sigma$
to $+1.7 \sigma$.

We also considered the Galactic dust correction alone and found that it accounts for $\sim$90\% of the
total change in the expected Galactic quadrupole moment, so it dominates over the correction for the locations
of ground telescopes.

%-----------------------------------------------
\subsection{Constraining the fraction of a possible Galactic population}
\label{sec:galfrac}
%-----------------------------------------------
The Galactic quadrupole moment $(\sin^2{b}-1/3)$ of unknown-$z$ events is marginally
deviating from isotropy: this holds both for the FREDs (S2) and,
to somewhat lower extent for the non-FREDs (SnFnZ).
This would imply an excess of events at low Galactic latitudes.
Taken at face value, when one overlooks the impact of the observational biases discussed in
Section~\ref{sec:bias} and plainly assumes that this is mostly due to the presence of a Galactic
population as T12 did, one can estimate its fraction within the sample of unknown-$z$ FREDs.

We did so by generating 100 synthetic samples with as many events as in S2, a fraction of which
was obtained from an isotropic distribution, and the remaining one from a random selection of a sample
of 334 known Galactic sources from the BAT catalogue by \citet{Cusumano10}.\footnote{We made use of
known sources just to mimic in a credible way what the distribution of a Galactic population
should look like. Clearly, such possible unrecognised Galactic sources must obviously be different
from the known ones.}
Both populations were selected compatibly with the exposure map.

By varying every time the fraction of the Galactic component, we ended up with the best match between
observations and simulations. This turned into an estimated fraction of Galactic intruders in S2
of $f=(27\pm10)$\% ($1 \sigma$), similar to T12's conclusions.
In practice, given that the evidence for anisotropy is only marginal, we provided a 90\%-confidence
upper limit of $f<40$\% by demanding that at least $90$\% of the synthetic samples had a non-zero
quadrupole moment at $>$$3 \sigma$.

However, neglecting the observational biases has the effect of overestimating the weight of the
possible Galactic population. We therefore repeated the same test by accounting for those biases.
Following the procedure of Section~\ref{sec:bias} we came up with a more realistic estimate of
$f=(19\pm11)$\% with a $90$\% upper limit of $f<34$\%.

%%%%%%%%%%%%%%%%%%%%%%%%%%%%%%%%%%%%%%%%%%%%%%%%%%%%%%%%%%%
\section{Discussion}
\label{sec:disc}
%%%%%%%%%%%%%%%%%%%%%%%%%%%%%%%%%%%%%%%%%%%%%%%%%%%%%%%%%%%
We found no evidence for a significant dipole moment in the different FRED samples observed by
{\em Swift}.
However, we point out that this does not clash with T12's claim: in contrast with what they assert,
$\cos{b}$ is by no means a dipole, since a dipole moment is the cosine of the angle between a generic
direction and a fixed one in the sky, not a plane.
Thus, the tension is just in the claim but not in the results, which do agree.
In particular, significant dipole moment was found neither along the Galactic poles ($\sin{b}$),
nor towards the Galactic centre ($\cos{b}\cos{l}$).

In contrast, and in agreement with T12, we found that the Galactic quadrupole moment
$(\sin^2{b}-1/3)$ of unknown-$z$ events marginally deviates from isotropy, implying
an excess of events at low Galactic latitudes.

However, before giving credence to the possible existence of such a relatively abundant disguised
Galactic community, one must carefully evaluate the impact of observational biases other than the
non-uniform exposure map.
That not just FREDs, but all GRBs with unknown $z$ tend to cluster towards the Galactic plane, likely
suggests that the impact of the process of measuring $z$ might be far from negligible.
This possibility appears to be corroborated by the absence of non-zero quadrupole moments in the entire
sample of BAT events (STot), in spite of its having the best statistical sensitivity.
Low-Galactic--latitude GRBs are notoriously disfavoured in their optical counterpart detection and
consequent $z$ measurement because of the strong Galactic dust extinction.
Moreover, our result that unknown-$z$ events tend to cluster around the celestial poles
($|\delta|>60^\circ$), where ground-based follow-up observations are less probable \citep{Fynbo09},
lends further support to the $z$-related observational bias explanation.
In the light of these considerations, we modelled the impact of both effects in shaping
the unknown-$z$ sample S2 by preliminarily studying how the measurement of $z$ is hampered by
the amount of Galactic dust content for the entire sample of {\em Swift} GRBs. Furthermore,
we modelled the probability for a generic source at a given declination to be followed up
spectroscopically by estimating its visibility from the most frequently used ground facilities.
As a result, we found that the deviation from isotropy of the Galactic quadrupole moment
decreased from $-2.8$$\sigma$ to $-1.8$$\sigma$, thus deflating the body of evidence of a significant
contamination due to a disguised population of Galactic transients.

We delved deeper into such matter by estimating the contaminating fraction of a Galactic population
that would yield the same Galactic quadrupole moment as the one observed in the sample of unknown-$z$ FREDs.
When the observational biases are properly accounted for, we estimate $f=(19\pm11)$\% (1$\sigma$) with a
$90$\% upper limit of $f<34$\%.

To gain further insight, we examined the spectral and temporal properties of the $\gamma$--ray
prompt emission along with the X--ray afterglow of the S2 sample as a function of the Galactic
position and compared their distributions with those derived for a sample of cosmological GRBs
with measured $z$. Specifically, we estimated the rise and the decay times of the $\gamma$--ray profiles
as well as the photon index in the 15--150~keV range of the spectrum extracted over the $T_{90}$
interval. The X--ray afterglow properties were taken from the classifications provided by {\em Swift}-XRT
catalogues \citep{Evans09,Margutti13} and from the online Leicester
catalogue\footnote{\url{http://www.swift.ac.uk/}} for the recent GRBs.
For none of these observables we found evidence for a different distribution between S2 and cosmological GRBs.

Noteworthy is that the fraction of short-duration GRBs in S2 is less than 7\% (8/119),
i.e. far from enough to account for the deviation from isotropy that we observe.
Several investigations on past short and very short GRB catalogues brought evidence for anisotropy
\citep{Balazs98,Magliocchetti03,Vavrek08,Cline03,Cline05}.
However, the low fraction of short GRBs in our {\em Swift} samples rules them out as the main cause
for the marginally observed anisotropy.

%%%%%%%%%%%%%%%%%%%%%%%%%%%%%%%%%%%%%%%%%%%%%%%%%%%%%%%%%%%
\section{Conclusions}
\label{sec:conc}
%%%%%%%%%%%%%%%%%%%%%%%%%%%%%%%%%%%%%%%%%%%%%%%%%%%%%%%%%%%
Triggered by previous results, we searched for the possible presence of disguised Galactic transients
in the catalogue of GRBs detected by {\em Swift}-BAT up to September 2014. We focused on FRED GRBs, given
that their light curves are often observed in a broad range of Galactic high-energy transients.
The search was based on the statistical analysis of the sky distribution both with reference to our Galaxy
and in a reference-system--independent way.
While no deviation from isotropy was found in the dipole moment, we found marginal ($2.8 \sigma$) evidence
for a non-zero Galactic quadrupole moment for FREDs with unknown redshift $z$, suggesting a clustering
at low Galactic latitudes, in particular towards the celestial poles ($|\delta|>60^\circ$).
Should this possible clustering on the Galactic plane be due the presence of an unidentified Galactic population,
we constrain its fraction to $f=(27\pm10)$\%.

However, non-FRED GRBs with unknown $z$ also show a similar behaviour, even though with lower
statistical significance ($2.1 \sigma$). If one drops the unknown-$z$ condition, such evidence disappears.
It is known that there are observational biases that disfavour the measurement of $z$: the large dust extinction
towards the Galactic plane along with the lower probability of ground-based observations of polar ($|\delta|>70^\circ$)
events, can contribute to enhance the degree of anisotropy.
Once such biases are accounted for, the evidence for a clustering near the Galactic plane decreases from
$\sim3$$\sigma$ to $\sim2$$\sigma$, thus weakening the claim for a Galactic contamination.
Furthermore, comparing the properties of the $\gamma$--ray prompt emission as well as of the X--ray
afterglow with those of GRBs with measured $z$ did not reveal any statistical difference.
Hence, our results suggest marginal evidence for the presence of an unrecognised Galactic population among
the GRBs with unknown redshift.

In conclusion, when $z$-measurement--related biases are accounted for, the fraction of a possible
disguised Galactic population of high-energy transients among {\em Swift}-BAT GRBs with unknown redshift
is estimated $f=(19\pm11)$\%, with a 90\%-confidence upper limit of $f<34$\%.

\begin{acknowledgements}
  We are grateful to the referee for comments which improved the paper.
  We thank G.~Cusumano, S.~Dichiara, F.~Frontera, C.~Koen, P.~Rosati,
  for useful discussions. PRIN MIUR project on ``Gamma Ray Bursts: from
  progenitors to physics of the prompt emission process'', P.~I. F. Frontera
  (Prot. 2009 ERC3HT) is acknowledged. This work made use of data supplied
  by the UK {\em Swift} Science Data Centre at the University of Leicester.
\end{acknowledgements}

%%%%%%%%%%%%%%%%%%%%%%%%%%%%%%%%%
% REFERENCES
%%%%%%%%%%%%%%%%%%%%%%%%%%%%%%%%%


\begin{thebibliography}{32}
\expandafter\ifx\csname natexlab\endcsname\relax\def\natexlab#1{#1}\fi

\bibitem[{{Balazs} {et~al.}(1998){Balazs}, {Meszaros}, \& {Horvath}}]{Balazs98}
{Balazs}, L.~G., {Meszaros}, A., \& {Horvath}, I. 1998, A\&A, 339, 1

\bibitem[{{Barthelmy} {et~al.}(2005){Barthelmy}, {Barbier}, {Cummings},
  {Fenimore}, {Gehrels}, {Hullinger}, {Krimm}, {Markwardt}, {Palmer},
  {Parsons}, {Sato}, {Suzuki}, {Takahashi}, {Tashiro}, \&
  {Tueller}}]{Barthelmy05}
{Barthelmy}, S.~D., {Barbier}, L.~M., {Cummings}, J.~R., {et~al.} 2005, Space
  Sci. Rev., 120, 143

\bibitem[{{Bernui} {et~al.}(2008){Bernui}, {Ferreira}, \&
  {Wuensche}}]{Bernui08}
{Bernui}, A., {Ferreira}, I.~S., \& {Wuensche}, C.~A. 2008, ApJ, 673, 968

\bibitem[{{Briggs}(1993)}]{Briggs93}
{Briggs}, M.~S. 1993, ApJ, 407, 126

\bibitem[{{Briggs} {et~al.}(1996){Briggs}, {Paciesas}, {Pendleton}, {Meegan},
  {Fishman}, {Horack}, {Brock}, {Kouveliotou}, {Hartmann}, \&
  {Hakkila}}]{Briggs96}
{Briggs}, M.~S., {Paciesas}, W.~S., {Pendleton}, G.~N., {et~al.} 1996, ApJ,
  459, 40

\bibitem[{{Burrows} {et~al.}(2005){Burrows}, {Hill}, {Nousek}, {Kennea},
  {Wells}, {Osborne}, {Abbey}, {Beardmore}, {Mukerjee}, {Short}, {Chincarini},
  {Campana}, {Citterio}, {Moretti}, {Pagani}, {Tagliaferri}, {Giommi},
  {Capalbi}, {Tamburelli}, {Angelini}, {Cusumano}, {Br{\"a}uninger}, {Burkert},
  \& {Hartner}}]{Burrows05}
{Burrows}, D.~N., {Hill}, J.~E., {Nousek}, J.~A., {et~al.} 2005, Space Sci.
  Rev., 120, 165

\bibitem[{{Castro-Tirado} {et~al.}(2008){Castro-Tirado}, {de Ugarte Postigo},
  {Gorosabel}, {Jel{\'{\i}}nek}, {Fatkhullin}, {Sokolov}, {Ferrero}, {Kann},
  {Klose}, {Sluse}, {Bremer}, {Winters}, {Nuernberger},
  {P{\'e}rez-Ram{\'{\i}}rez}, {Guerrero}, {French}, {Melady}, {Hanlon},
  {McBreen}, {Leventis}, {Markoff}, {Leon}, {Kraus}, {Aceituno}, {Cunniffe},
  {Kub{\'a}nek}, {V{\'{\i}}tek}, {Schulze}, {Wilson}, {Hudec}, {Durant},
  {Gonz{\'a}lez-P{\'e}rez}, {Shahbaz}, {Guziy}, {Pandey}, {Pavlenko}, {Sonbas},
  {Trushkin}, {Bursov}, {Nizhelskij}, {S{\'a}nchez-Fern{\'a}ndez}, \&
  {Sabau-Graziati}}]{CastroTirado08}
{Castro-Tirado}, A.~J., {de Ugarte Postigo}, A., {Gorosabel}, J., {et~al.}
  2008, Nature, 455, 506

\bibitem[{{Cline} {et~al.}(2005){Cline}, {Czerny}, {Matthey}, {Janiuk}, \&
  {Otwinowski}}]{Cline05}
{Cline}, D.~B., {Czerny}, B., {Matthey}, C., {Janiuk}, A., \& {Otwinowski}, S.
  2005, ApJ, 633, L73

\bibitem[{{Cline} {et~al.}(2003){Cline}, {Matthey}, \& {Otwinowski}}]{Cline03}
{Cline}, D.~B., {Matthey}, C., \& {Otwinowski}, S. 2003, Astroparticle Physics,
  18, 531

\bibitem[{{Cucchiara} {et~al.}(2011){Cucchiara}, {Levan}, {Fox}, {Tanvir},
  {Ukwatta}, {Berger}, {Kr{\"u}hler}, {K{\"u}pc{\"u} Yolda{\c s}}, {Wu},
  {Toma}, {Greiner}, {Olivares}, {Rowlinson}, {Amati}, {Sakamoto}, {Roth},
  {Stephens}, {Fritz}, {Fynbo}, {Hjorth}, {Malesani}, {Jakobsson}, {Wiersema},
  {O'Brien}, {Soderberg}, {Foley}, {Fruchter}, {Rhoads}, {Rutledge}, {Schmidt},
  {Dopita}, {Podsiadlowski}, {Willingale}, {Wolf}, {Kulkarni}, \&
  {D'Avanzo}}]{Cucchiara11b}
{Cucchiara}, A., {Levan}, A.~J., {Fox}, D.~B., {et~al.} 2011, ApJ, 736, 7

\bibitem[{{Cusumano} {et~al.}(2010){Cusumano}, {La Parola}, {Segreto},
  {Ferrigno}, {Maselli}, {Sbarufatti}, {Romano}, {Chincarini}, {Giommi},
  {Masetti}, {Moretti}, {Parisi}, \& {Tagliaferri}}]{Cusumano10}
{Cusumano}, G., {La Parola}, V., {Segreto}, A., {et~al.} 2010, A\&A, 524, A64

\bibitem[{{Evans} {et~al.}(2009){Evans}, {Beardmore}, {Page}, {Osborne},
  {O'Brien}, {Willingale}, {Starling}, {Burrows}, {Godet}, {Vetere}, {Racusin},
  {Goad}, {Wiersema}, {Angelini}, {Capalbi}, {Chincarini}, {Gehrels}, {Kennea},
  {Margutti}, {Morris}, {Mountford}, {Pagani}, {Perri}, {Romano}, \&
  {Tanvir}}]{Evans09}
{Evans}, P.~A., {Beardmore}, A.~P., {Page}, K.~L., {et~al.} 2009, MNRAS, 397,
  1177

\bibitem[{{Fynbo} {et~al.}(2009){Fynbo}, {Jakobsson}, {Prochaska}, {Malesani},
  {Ledoux}, {de Ugarte Postigo}, {Nardini}, {Vreeswijk}, {Wiersema}, {Hjorth},
  {Sollerman}, {Chen}, {Th{\"o}ne}, {Bj{\"o}rnsson}, {Bloom}, {Castro-Tirado},
  {Christensen}, {De Cia}, {Fruchter}, {Gorosabel}, {Graham}, {Jaunsen},
  {Jensen}, {Kann}, {Kouveliotou}, {Levan}, {Maund}, {Masetti},
  {Milvang-Jensen}, {Palazzi}, {Perley}, {Pian}, {Rol}, {Schady}, {Starling},
  {Tanvir}, {Watson}, {Xu}, {Augusteijn}, {Grundahl}, {Telting}, \&
  {Quirion}}]{Fynbo09}
{Fynbo}, J.~P.~U., {Jakobsson}, P., {Prochaska}, J.~X., {et~al.} 2009, ApJS,
  185, 526

\bibitem[{{Guidorzi}(2015)}]{Guidorzi15a}
{Guidorzi}, C. 2015, Astronomy and Computing, 10, 54

\bibitem[{{Hartmann} \& {Blumenthal}(1989)}]{HartmannBlumenthal89}
{Hartmann}, D. \& {Blumenthal}, G.~R. 1989, ApJ, 342, 521

\bibitem[{{Hartmann} \& {Epstein}(1989)}]{HartmannEpstein89}
{Hartmann}, D. \& {Epstein}, R.~I. 1989, ApJ, 346, 960

\bibitem[{{Kasliwal} {et~al.}(2008){Kasliwal}, {Cenko}, {Kulkarni}, {Cameron},
  {Nakar}, {Ofek}, {Rau}, {Soderberg}, {Campana}, {Bloom}, {Perley}, {Pollack},
  {Barthelmy}, {Cummings}, {Gehrels}, {Krimm}, {Markwardt}, {Sato}, {Chandra},
  {Frail}, {Fox}, {Price}, {Berger}, {Grebenev}, {Krivonos}, \&
  {Sunyaev}}]{Kasliwal08}
{Kasliwal}, M.~M., {Cenko}, S.~B., {Kulkarni}, S.~R., {et~al.} 2008, ApJ, 678,
  1127

\bibitem[{{Lewin} {et~al.}(1993){Lewin}, {van Paradijs}, \&
  {Taam}}]{Lewin93rev}
{Lewin}, W.~H.~G., {van Paradijs}, J., \& {Taam}, R.~E. 1993, SSRv, 62, 223

\bibitem[{{Magliocchetti} {et~al.}(2003){Magliocchetti}, {Ghirlanda}, \&
  {Celotti}}]{Magliocchetti03}
{Magliocchetti}, M., {Ghirlanda}, G., \& {Celotti}, A. 2003, MNRAS, 343, 255

\bibitem[{{Margutti} {et~al.}(2013){Margutti}, {Zaninoni}, {Bernardini},
  {Chincarini}, {Pasotti}, {Guidorzi}, {Angelini}, {Burrows}, {Capalbi},
  {Evans}, {Gehrels}, {Kennea}, {Mangano}, {Moretti}, {Nousek}, {Osborne},
  {Page}, {Perri}, {Racusin}, {Romano}, {Sbarufatti}, {Stafford}, \&
  {Stamatikos}}]{Margutti13}
{Margutti}, R., {Zaninoni}, E., {Bernardini}, M.~G., {et~al.} 2013, MNRAS, 428,
  729

\bibitem[{{Mereghetti}(2008)}]{Mereghetti08rev}
{Mereghetti}, S. 2008, A\&ARv, 15, 225

\bibitem[{{M{\'e}sz{\'a}ros} \& {Gehrels}(2012)}]{MeszarosGehrels12}
{M{\'e}sz{\'a}ros}, P. \& {Gehrels}, N. 2012, Research in Astronomy and
  Astrophysics, 12, 1139

\bibitem[{{Norris} {et~al.}(1996){Norris}, {Nemiroff}, {Bonnell}, {Scargle},
  {Kouveliotou}, {Paciesas}, {Meegan}, \& {Fishman}}]{Norris96}
{Norris}, J.~P., {Nemiroff}, R.~J., {Bonnell}, J.~T., {et~al.} 1996, ApJ, 459,
  393

\bibitem[{{Pietka} {et~al.}(2015){Pietka}, {Fender}, \& {Keane}}]{Pietka15}
{Pietka}, M., {Fender}, R.~P., \& {Keane}, E.~F. 2015, MNRAS, 446, 3687

\bibitem[{{Remillard} \& {McClintock}(2006)}]{Remillard06rev}
{Remillard}, R.~A. \& {McClintock}, J.~E. 2006, ARAA, 44, 49

\bibitem[{{Schlafly} \& {Finkbeiner}(2011)}]{Schlafly11}
{Schlafly}, E.~F. \& {Finkbeiner}, D.~P. 2011, ApJ, 737, 103

\bibitem[{{Segreto} {et~al.}(2010){Segreto}, {Cusumano}, {Ferrigno}, {La
  Parola}, {Mangano}, {Mineo}, \& {Romano}}]{Segreto10}
{Segreto}, A., {Cusumano}, G., {Ferrigno}, C., {et~al.} 2010, A\&A, 510, A47

\bibitem[{{Stefanescu} {et~al.}(2008){Stefanescu}, {Kanbach}, {S{\l}owikowska},
  {Greiner}, {McBreen}, \& {Sala}}]{Stefanescu08}
{Stefanescu}, A., {Kanbach}, G., {S{\l}owikowska}, A., {et~al.} 2008, Nature,
  455, 503

\bibitem[{{Tanvir} {et~al.}(2005){Tanvir}, {Chapman}, {Levan}, \&
  {Priddey}}]{Tanvir05}
{Tanvir}, N.~R., {Chapman}, R., {Levan}, A.~J., \& {Priddey}, R.~S. 2005,
  Nature, 438, 991

\bibitem[{{Tegmark}(1996)}]{Tegmark96}
{Tegmark}, M. 1996, ApJ, 470, L81

\bibitem[{{Tello} {et~al.}(2012){Tello}, {Castro-Tirado}, {Gorosabel},
  {P{\'e}rez-Ram{\'{\i}}rez}, {Guziy}, {S{\'a}nchez-Ram{\'{\i}}rez},
  {Jel{\'{\i}}nek}, {Veres}, \& {Bagoly}}]{Tello12}
{Tello}, J.~C., {Castro-Tirado}, A.~J., {Gorosabel}, J., {et~al.} 2012, A\&A,
  548, L7

\bibitem[{{Vavrek} {et~al.}(2008){Vavrek}, {Bal{\'a}zs}, {M{\'e}sz{\'a}ros},
  {Horv{\'a}th}, \& {Bagoly}}]{Vavrek08}
{Vavrek}, R., {Bal{\'a}zs}, L.~G., {M{\'e}sz{\'a}ros}, A., {Horv{\'a}th}, I.,
  \& {Bagoly}, Z. 2008, MNRAS, 391, 1741

\end{thebibliography}
\end{document}